\documentclass[10pt,twocolumn,letterpaper]{article}

\usepackage[utf8]{inputenc}
\usepackage[T1]{fontenc}
\usepackage[margin=0.75in,top=0.75in,bottom=1in]{geometry}
\usepackage{graphicx}
\usepackage{hyperref}
\usepackage{url}
\usepackage[english]{babel}
\usepackage{adjustbox}
\usepackage{tabularx}
\usepackage{booktabs}
\usepackage{makecell}
\usepackage{subcaption}
\usepackage{amsmath}
\usepackage{amssymb}
\usepackage{titlesec}
\usepackage{abstract}
\usepackage{parskip}

\titleformat{\section}{\normalfont\bfseries\scshape}{\Roman{section}.}{0.5em}{}
\titleformat{\subsection}{\normalfont\itshape}{\Alph{subsection}.}{0.5em}{}
\titlespacing*{\section}{0pt}{8pt}{4pt}
\titlespacing*{\subsection}{0pt}{5pt}{2pt}

\makeatletter
\renewcommand{\@biblabel}[1]{[#1]}
\renewcommand{\thebibliography}[1]{%
  \section*{\textbf{References}}
  \small
  \list{\@biblabel{\@arabic\c@enumiv}}{%
    \settowidth\labelwidth{\@biblabel{#1}}%
    \leftmargin\labelwidth
    \advance\leftmargin\labelsep
    \usecounter{enumiv}%
    \let\p@enumiv\@empty
    \renewcommand\theenumiv{\@arabic\c@enumiv}}%
  \sloppy\clubpenalty4000\widowpenalty4000%
  \sfcode`\.\@m}
\makeatother

\title{\LARGE \textbf{Coexisting Tempo Traditions in Beethoven's Piano and Cello Sonatas: A K-means Clustering Analysis of Recorded Performances, 1930--2012}}

\author{Dr Ignasi Sole \quad \texttt{ignasiphd@gmail.com} \quad \today}

\begin{document}

\maketitle
\thispagestyle{empty}
\pagestyle{empty}


\begin{abstract}
Empirical studies of recorded performance have conventionally modelled tempo change as a unidirectional historical process, fitting linear regression lines to tempo data plotted against recording year. This paper argues that such single-regression approaches impose a false narrative of uniform stylistic evolution on what is, in fact, a plurality of coexisting interpretive traditions. Applying $k$-means clustering ($k=3$, with $k$-means$^{++}$ initialisation and 100 random restarts) to bar-level beats-per-minute data from over one hundred movement-level recordings of Beethoven's five piano and cello sonatas (Op.~5 Nos.~1 and~2; Op.~69; Op.~102 Nos.~1 and~2) spanning 1930--2012, this study reveals that every movement supports at least two, and usually three, discrete tempo traditions (labelled slow, mid-range, and fast), whose internal regression slopes are negligible (\(R^{2} \leq 0.25\) in all but one case), demonstrating that each tradition is independently stable across eight decades. The mid-range cluster dominates in all movements, typically comprising 55--70\% of recordings. A slow cluster is absent from fast-character movements (Op.~5 Rondos, Op.~69 Scherzo), reflecting a widely shared rhetorical consensus about their character. The single case of statistically significant intra-cluster drift (the Op.~102 No.~1 Allegro con brio (\(R^{2}=0.246\), \(p=0.013\))) indicates a moderate mid-range deceleration of approximately 3.2~BPM across the study period. No correlation is found between cluster membership and performers' generational, national, or pedagogical backgrounds, indicating that tempo tradition is a product of individual interpretive choice rather than collective cultural inheritance. The paper proposes an ecological model of stylistic change (coexisting traditions shifting in relative prevalence rather than a single tradition evolving), and argues that this reframing has broad implications for how empirical performance studies interpret corpus-level tempo data.
\end{abstract}


\section{Introduction}

Since Bowen's 1996 study demonstrated that tempo, duration, and expressive flexibility are measurable across large corpora of recordings~\cite{c1}, empirical performance analysis has increasingly relied on regression-based modelling to characterise historical trends. The standard methodology is familiar: compute a representative tempo value per recording (typically the movement mean), plot it against recording year, and fit a linear regression line. The slope of that line is then read as evidence of a historical tendency; an acceleration, a deceleration, or relative stability.

This methodology is appropriate when the underlying population of performances genuinely follows a single distributional trajectory. It is inappropriate (and potentially misleading) when that population is in fact composed of several discrete subpopulations pursuing different tempo centres simultaneously. In that case, a single regression line does not describe any performer's actual behaviour; it describes a fictional average performer who exists in no recording and belongs to no tradition.

The hypothesis motivating this paper is that the tempo distributions of Beethoven's piano and cello sonata recordings are of the second kind. At any given moment between 1930 and 2012, performers pursuing slow, moderate, and fast interpretations of the same movement coexisted. Historical change, where it occurred, was not a matter of all performers shifting uniformly in one direction, but of the proportional balance among these coexisting traditions shifting over time. Identifying and characterising those traditions separately is a more accurate description of the historical record than any single regression line can provide.

To test this hypothesis, this paper applies $k$-means unsupervised clustering to bar-level BPM data from a corpus of over one hundred movement-level recordings spanning eight decades. The results reveal a consistent tripartite structure in most movements, with each cluster exhibiting internal stability far greater than the full-corpus regression would suggest. The paper then asks what drives cluster membership; and finds that it is not generation, national school, or pedagogical lineage, but individual interpretive choice.

The paper proceeds as follows. Section~II reviews the literature on tempo analysis and historical performance trends, identifying the assumptions embedded in regression-based methods and their consequences. Section~III presents the clustering methodology in full. Section~IV reports the results movement by movement. Section~V synthesises the findings into a cross-movement and cross-sonata comparison. Section~VI develops the ecological model of stylistic change and discusses implications. Section~VII concludes.


\section{Background}

\subsection{Regression-Based Tempo Analysis and Its Limits}

The regression approach to historical tempo analysis is well established. Bowen's scatter plots of performance duration against average tempo revealed that conductors maintain consistent internal structural proportions even when average tempi differ significantly, and that overall duration is shaped less by average tempo than by internal tempo modulation~\cite{c1}. Cook's analysis of over 1,700 recordings of Chopin's Mazurkas used tempo curves and correlation networks to propose performance genealogies based on stylistic similarity, noting that teacher-pupil lineages could sometimes be traced in the resulting clusters~\cite{c2}. Leech-Wilkinson's empirical studies of portamento and timing in violin and cello recordings across multiple repertoires documented systematic decline in expressive devices across the twentieth century, attributing this to the influence of recording technology and changing aesthetic norms~\cite{c3}. Philip's broader survey of early twentieth-century recordings reached similar conclusions about the general direction of stylistic change~\cite{c4}.

What these studies share (and what this paper questions), is an implicit assumption of unidirectionality: that there is a single prevailing tendency at each historical moment, deviations from which are treated as individual eccentricities rather than as evidence of a genuinely different tradition. The regression line is both the method and the assumption: by fitting one line, the analyst commits to the existence of one trend.

This assumption has been questioned implicitly by findings within the literature itself. Cook's correlation networks for the Chopin Mazurkas suggested that multiple performance genealogies coexisted rather than converging toward a single style~\cite{c2}. Leech-Wilkinson acknowledged that ``old-fashioned'' performance features persisted long after they had allegedly disappeared from mainstream practice~\cite{c3}. Repp, in his analysis of timing patterns across nineteen pianists performing a Beethoven Minuet, found no universal pattern, only a distribution of individual approaches that resisted reduction to a single norm~\cite{c5}. These findings are consistent with a model in which multiple traditions coexist, but none of these studies had the methodological framework to identify and characterise those traditions as discrete entities.

\subsection{Unsupervised Clustering in Music Analysis}

Unsupervised clustering methods have been applied to music in a variety of contexts, primarily in Music Information Retrieval research. Clustering has been used to segment audio signals, group similar melodic patterns, and classify timbral profiles~\cite{c6}. Its application to historical performance data as a historiographical tool (not to classify audio features but to identify interpretive traditions in a corpus of tempo measurements) is, without direct precedent.

The $k$-means algorithm, the most widely used unsupervised clustering method, partitions $n$ observations into $k$ clusters by minimising the total within-cluster sum of squared Euclidean distances from each point to its cluster centroid~\cite{c7}. Its primary limitation (sensitivity to initial centroid placement and susceptibility to local minima), is addressed by $k$-means$^{++}$ initialisation and multiple random restarts. Its primary advantage, for the purpose of this study, is interpretability: the resulting clusters can be named, described, and compared in terms that are musically meaningful (slow, mid-range, fast), and the centroid of each cluster represents a prototypical tempo approach that can be contextualised within the historical and stylistic record.

\subsection{The Ecology of Performance Traditions}

This paper proposes an ecological frame for understanding the results. In an ecological system, multiple species occupy a shared environment, competing for resources and fluctuating in relative population without any necessarily driving the others to extinction. Stylistic traditions in performance practice work analogously. Slow-tempo, mid-range, and fast-tempo approaches to a given Beethoven movement coexist at any historical moment, pursued by different performers for different reasons. Historical change is not the replacement of one tradition by another but a shift in their relative populations: the fast cluster may become more densely populated after 1970, the slow cluster may thin, but neither disappears entirely.

This ecological model contrasts with the standard evolutionary model implicit in regression-based studies, where stylistic change is conceptualised as movement along a single dimension from an older to a newer norm. The ecological model makes no such assumption: it allows for simultaneous plurality, for the persistence of minority traditions alongside majority ones, and for the possibility that what looks like a historical trend at the aggregate level is actually a change in the proportional balance among stable sub-traditions.

Pati, Lerch, Arthur, and Gururani's interdisciplinary review of music performance analysis recommends comparing multiple recordings of a single work across performers and career stages to approach ``the nature of the performance parameter''~\cite{c8}. This paper operationalises that recommendation at the level of the entire corpus.


\section{Methodology}

\subsection{Corpus and Data}

The corpus consists of recordings of the second and third movements of Beethoven's five piano and cello sonatas: Op.~5 No.~1, Op.~5 No.~2, Op.~69, Op.~102 No.~1, and Op.~102 No.~2. Recordings date from 1930 to 2012. For each movement, between 18 and 22 recordings were analysed, yielding a total of over one hundred movement-level data points. Performers include Pau Casals, Emanuel Feuermann, Pierre Fournier, Jacqueline du~Pr\'e, Mstislav Rostropovich, Yo-Yo Ma, Pieter Wispelwey, Anner Bylsma, Mischa Maisky, Janos Starker, Zara Nelsova, Daniel Shafran, Paul Tortelier, Steven Isserlis, M\'ikl\'os Per\'enyi, Maria Kliegel, Julius Klengel, and others.

Tempo data were collected using the manual bar-by-bar stopwatch protocol described in Sole~\cite{c9}, which was developed in response to the systematic failure of automated beat-detection tools on polyphonic duo recordings. For each movement and each recording, the primary feature used for clustering was the mean BPM across all bars of the movement. For movements where internal tempo flexibility was analytically pertinent, the coefficient of variation (CV) of the bar-level BPM values was included as a second feature.

\subsection{Feature Engineering and Standardisation}

For fast movements (the Rondo finales of Op.~5 Nos.~1 and~2; the Scherzo and Allegro vivace of Op.~69; the Allegro con brio of Op.~102 No.~1; the Allegro of Op.~102 No.~2), the primary clustering feature was the movement mean BPM. For slow introductions (the Adagio cantabile of Op.~69; the Adagio of Op.~102 No.~1; the Adagio of Op.~102 No.~2), the mean BPM was supplemented by the CV of instantaneous tempo where this added discriminatory power between clusters.

All features were $z$-standardised prior to clustering: each feature was transformed to zero mean and unit variance by subtracting the sample mean and dividing by the sample standard deviation. This ensures that features measured on different scales contribute equally to the Euclidean distance used by the $k$-means algorithm, preventing mean BPM from dominating cluster membership at the expense of the CV where both are included.

\subsection{K-means Procedure}

The $k$-means algorithm was applied with $k=3$ clusters to each movement's standardised feature set. The target of three clusters was motivated by the theoretical expectation of three coexisting interpretive traditions (slow, mid-range, fast), a prior that is musically grounded rather than statistically arbitrary: listeners and performers have long described performances of classical works in these three-way terms. Formal cluster validity tests; the elbow method (within-cluster sum of squares plotted against $k$) and silhouette scores (a measure of how much more similar each point is to its own cluster than to the nearest alternative), were applied to confirm that $k=3$ was supported by the data for each movement. In cases where validity tests indicated fewer than three meaningful clusters (specifically, Op.~5 No.~2 Rondo, Op.~69 Scherzo), the empty cluster was reported as such rather than forcing a three-way partition.

To guard against poor local minima, $k$-means$^{++}$ initialisation was used for all runs: initial centroids were chosen by sampling proportionally to their squared distance from previously chosen centroids, ensuring maximal initial spread across the data space~\cite{c10}. The algorithm was run 100 times with different random seeds, and the solution with the lowest total within-cluster inertia was retained. Clusters were then labelled slow, mid-range, and fast by ordering centroids from lowest to highest mean BPM.

\subsection{Intra-Cluster Regression}

To test whether each identified tradition was itself stable across time or showed its own directional drift, a separate linear regression of BPM against recording year was fitted within each cluster. The slope and $R^{2}$ of these intra-cluster regressions are the primary evidence for or against within-tradition temporal stability. A near-zero slope with low $R^{2}$ indicates that cluster membership, not calendar year, is the dominant predictor of a recording's tempo. A statistically significant slope within a cluster indicates that even the members of a given tradition have drifted over time, which is the more interesting and rarer finding.


\section{Results}

\subsection{Op.~5 No.~1, Rondo Allegro Vivace}

\begin{figure}[htbp]
  \centering
  \includegraphics[width=\columnwidth]{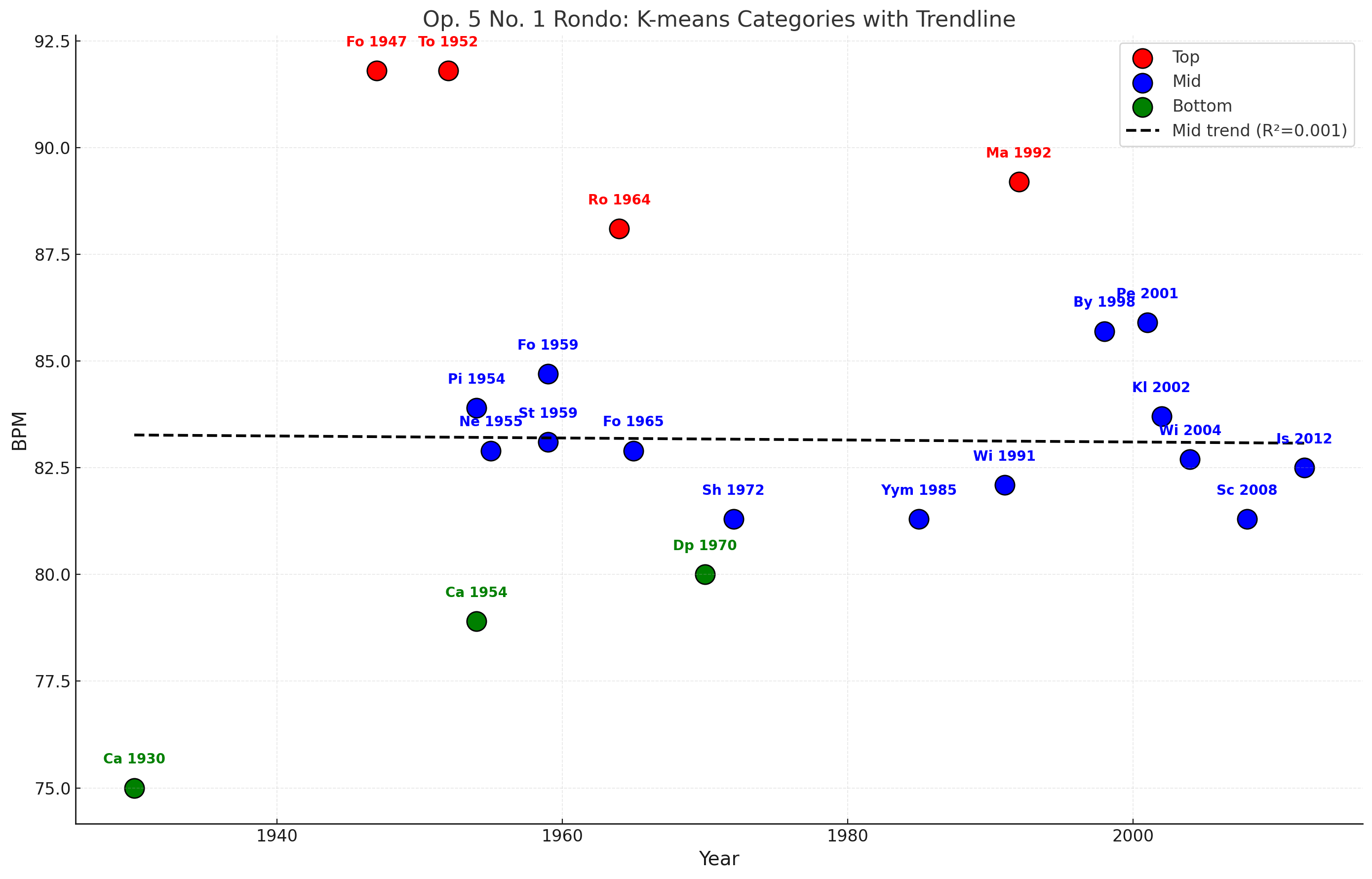}
  \caption{Scatter plot of average tempo by recording year for the Op.\,5 No.\,1 Rondo, colour-coded by cluster (slow/mid/fast). Dashed line shows linear regression on the mid-range cluster ($R^{2}=0.001$).}
  \label{fig:op5n1-rondo}
\end{figure}

Applying $k$-means ($k=3$) to the Rondo yields three fully populated clusters. The slow cluster (mean 78.0~BPM) comprises Casals (1930, 1954) and du~Pr\'e (1970), three recordings spanning 1930--1970 that share an expansive approach predating the mid-century consensus. The mid-range cluster (mean 83.1~BPM) is the largest, grouping thirteen recordings from Piatigorsky (1954) through Isserlis (2012) and representing the normative pace adopted by most cellists after mid-century. The fast cluster (mean 90.2~BPM) consists of four outliers: Fournier (1947--48), Tortelier (1952), Rostropovich (1964), and Maisky (1992).

A linear regression on the mid-range cluster returns a slope indistinguishable from zero ($R^{2}=0.001$), confirming that the dominant tradition has been internally stable across eight decades. The existence of the slow cluster exclusively in the 1930--1970 period and its subsequent absence are consistent with an ecological reading: this tradition became depopulated, not extinct, as mid-century pedagogy consolidated around a faster normative tempo.

\subsection{Op.~5 No.~2, Rondo Allegro}

\begin{figure}[htbp]
    \centering
  \includegraphics[width=\columnwidth]{Op_5_No_2_Rondo_Cluster.png}
    \caption{Scatter plot of average tempo by recording year for the Op.~5 No.~2 Rondo. Two populated clusters: mid-range (blue) and fast (red). No slow cluster emerged (\(R^{2}=0.142\) for mid-range).}
    \label{fig:op5n2-rondo}
\end{figure}

The Op.~5 No.~2 Rondo is the only movement in the corpus where $k$-means returns only two meaningful clusters: a mid-range group of fourteen recordings centred on 66.5~BPM, and a fast group of five recordings centred on 76.7~BPM. No slow cluster materialises, indicating that performers treat this Rondo as having a tighter rhetorical consensus than the first sonata's finale. The mid-range regression slope is modest ($R^{2}=0.142$), accounting for only 14.2\% of the intra-cluster tempo variance. The remaining 85.8\% reflects individual interpretive choice, not calendar year.

Comparing the two Op.~5 Rondos: the first supports three clusters with slow--mid and mid--fast gaps exceeding 5~BPM each; the second supports only two, with a 10~BPM gap between mid and fast but no slow tradition at all. This contrast suggests that performers hear the two Rondos as distinct rhetorical entities: the first's more varied internal structure invites a wider tempo palette, while the second's more consistently driven character discourages any expansive slow reading.

\subsection{Op.~69, Scherzo}

\begin{figure}[htbp]
    \centering
    \includegraphics[width=\columnwidth]{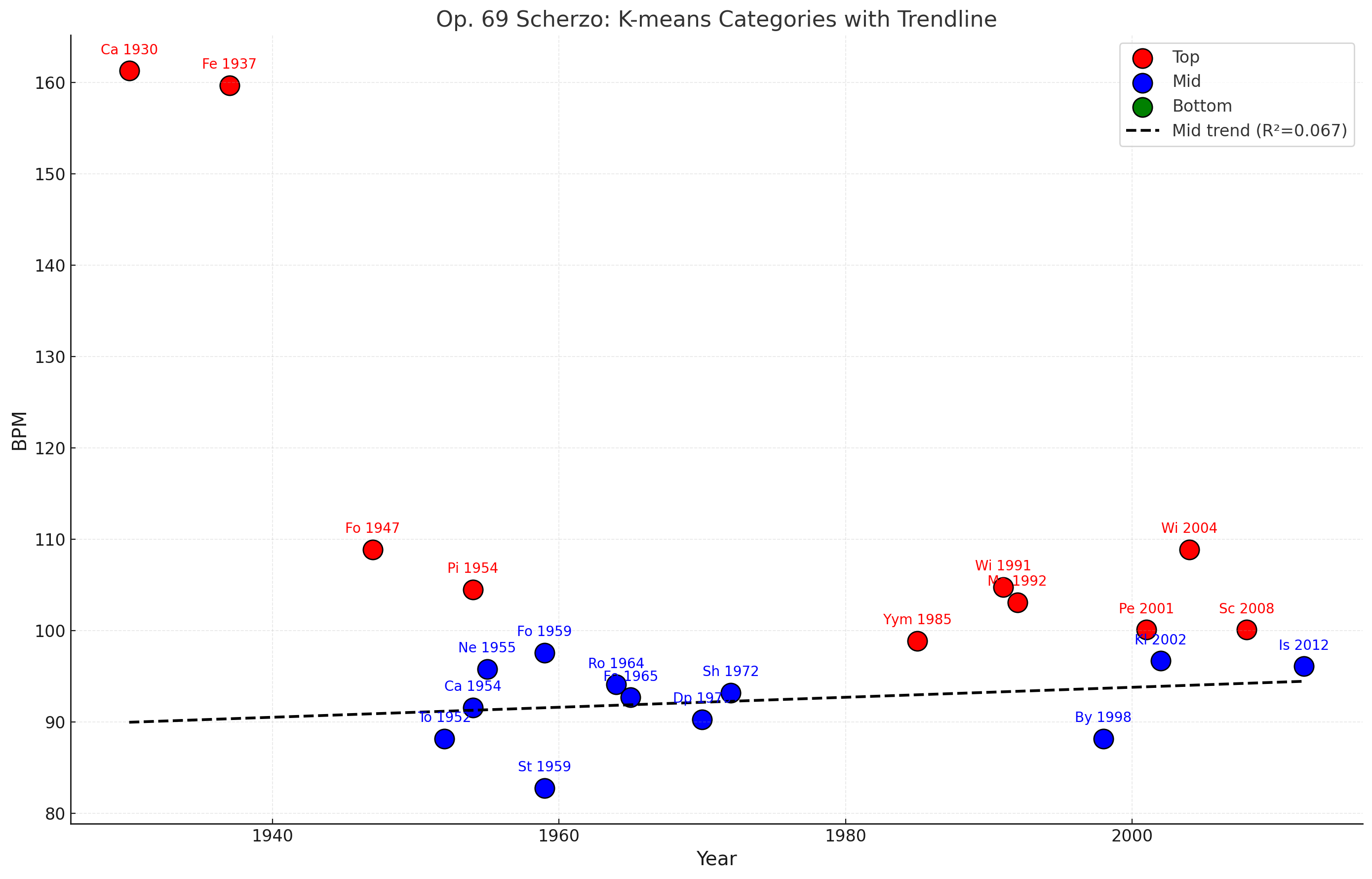}
    \caption{Scatter plot of average tempo by recording year for the Op.~69 Scherzo. Mid-range cluster (blue, mean 92.3~BPM) and fast cluster (red) are populated; no slow cluster emerged. Mid-range regression slope $\approx -0.013$~BPM~yr$^{-1}$ ($R^{2}=0.067$).}
    \label{fig:op69-scherzo}
\end{figure}

The Op.~69 Scherzo also yields two populated clusters and one empty slow category. The mid-range cluster (mean 92.3~BPM, SD 3.5~BPM) comprises fourteen recordings from Piatigorsky (1954) through Schott (2008), reflecting a collective preference for tempi between 88 and 98~BPM that emphasises clarity of articulation. The fast cluster is characterised by extreme outliers: Casals (1930, 161.3~BPM) and Feuermann (1937, 159.7~BPM) stand as the most remote, while Rostropovich (1964) and Wispelwey (2004) occupy an intermediate fast zone.

The mid-range intra-cluster regression yields a slope of $-0.013$~BPM~yr$^{-1}$ ($R^{2}=0.067$), indicating an almost imperceptible deceleration of approximately 1.3~BPM across the entire study period. The extreme early outliers are notable: they are not evidence of an early fast tradition that subsequently collapsed, but of isolated individual choices by two of the twentieth century's most idiosyncratic cellists. Both Casals and Feuermann were outliers in their own historical moment, not representatives of a slow-tempo tradition in the Op.~5 sense.

\subsection{Op.~69, Adagio Cantabile}

\begin{figure}[htbp]
    \centering
    \includegraphics[width=\columnwidth]{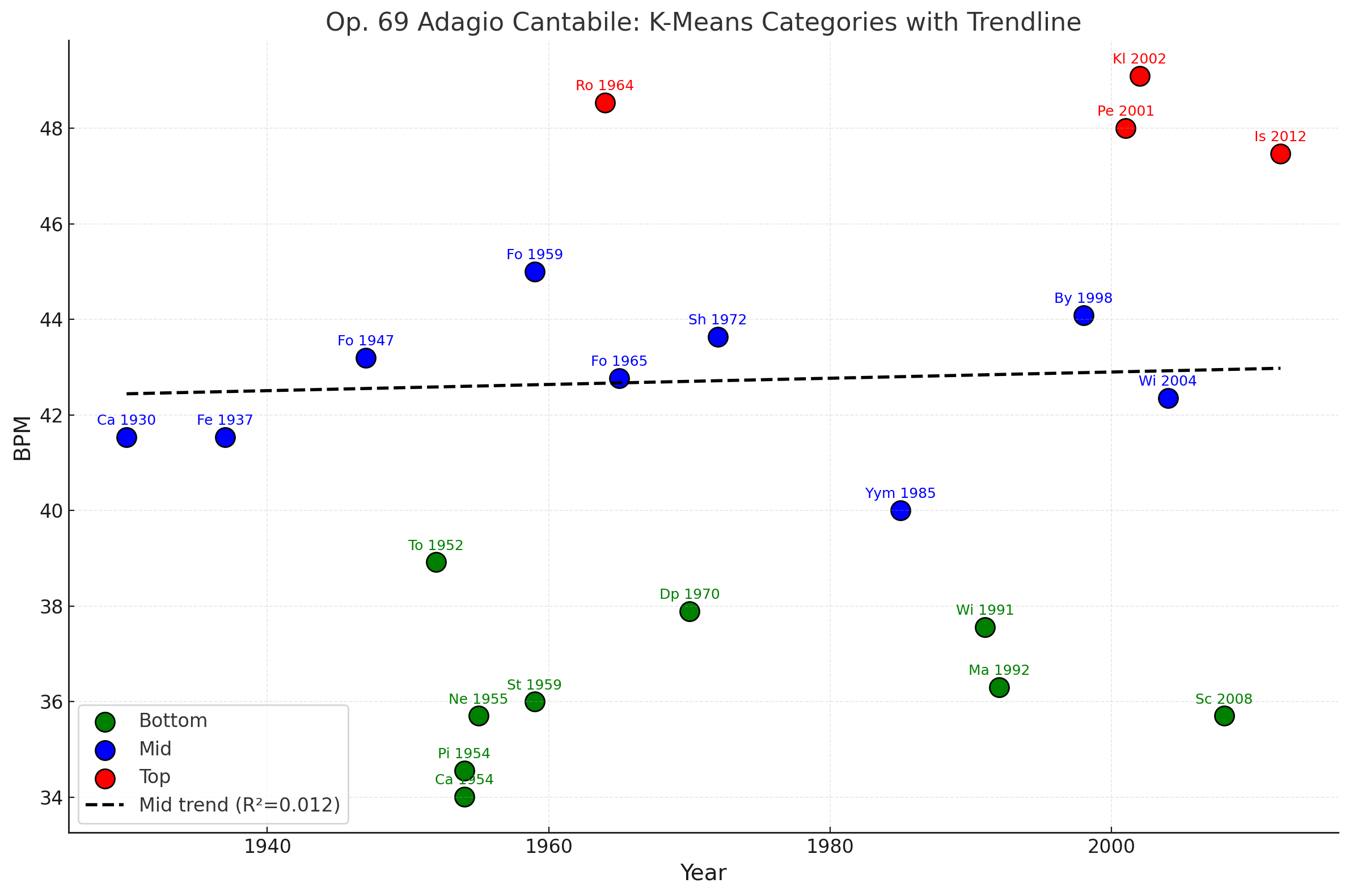}
    \caption{Scatter plot of average tempo by recording year for the Op.~69 Adagio cantabile introduction. Three clusters: slow (green, mean 36.3~BPM), mid-range (blue, mean 42.7~BPM), fast (red, mean 48.3~BPM). Mid-range regression slope $\approx +0.018$~BPM~yr$^{-1}$ ($R^{2}=0.012$).}
    \label{fig:op69-adagio}
\end{figure}

The 18-bar Adagio cantabile introduction to Op.~69 yields all three clusters fully populated, with a slow group centred on 36.3~BPM (Tortelier, Piatigorsky, Casals 1954, Nelsova, Starker, Ma 1992), a mid-range group averaging 42.7~BPM across ten recordings from Casals (1930) through Wispelwey (2004), and a fast group of four recordings averaging 48.3~BPM (Rostropovich, Per\'enyi, Kliegel, Isserlis). The gap between slow and fast centroids is approximately 12~BPM --- a wide spread for an 18-bar introduction.

The mid-range intra-cluster slope is positive ($+0.018$~BPM~yr$^{-1}$, $R^{2}=0.012$), indicating an imperceptible marginal acceleration, consistent with the interpretation that modern performers increasingly treat this introduction as a propulsive prelude rather than an extended moment of expressive stasis. Yet the slow cluster persists: its presence in the data shows that some performers throughout the study period continued to treat this opening as an opportunity for genuine expressive expansion.

\subsection{Op.~69, Allegro Vivace}

\begin{figure}[htbp]
    \centering
    \includegraphics[width=\columnwidth]{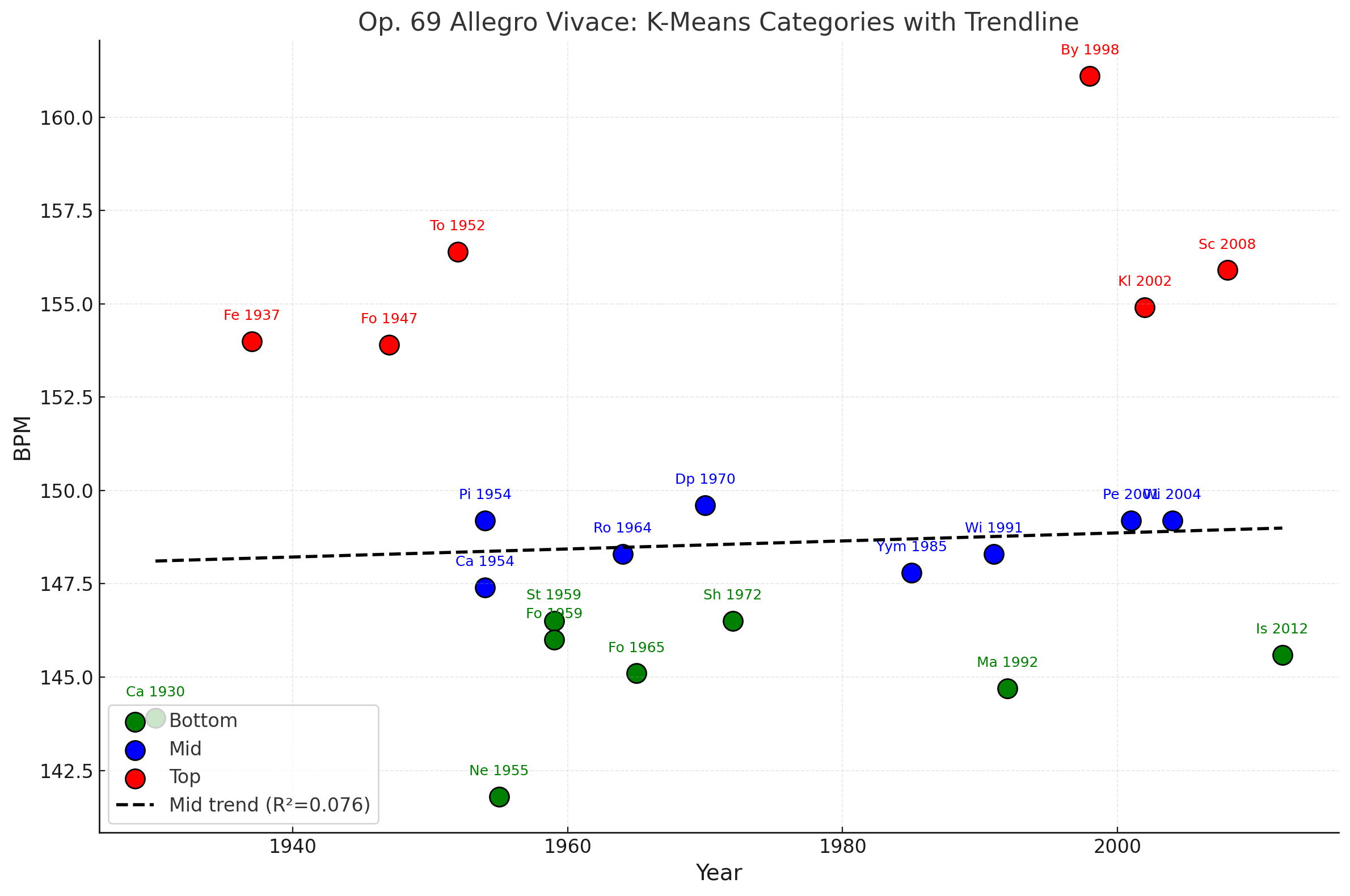}
    \caption{Scatter plot of average tempo by recording year for the Op.~69 Allegro vivace. Three clusters: slow (green, mean 145.0~BPM), mid-range (blue, mean 148.6~BPM), fast (red, mean 156.0~BPM). Mid-range regression slope $\approx -0.015$~BPM~yr$^{-1}$ ($R^{2}=0.076$).}
    \label{fig:op69-allegro}
\end{figure}

The Op.~69 Allegro vivace produces three clusters separated by remarkably narrow absolute margins: slow (mean 145.0~BPM, six recordings), mid-range (mean 148.6~BPM, seven recordings), and fast (mean 156.0~BPM, nine recordings). The slow--mid gap of 3.6~BPM and the mid--fast gap of 7.4~BPM are the smallest in the corpus, reflecting the fact that this movement's character permits little radical tempo divergence at the upper end: the tempo is already fast by definition, and the range within which it can be varied without becoming unmusicality is narrow.

The mid-range regression slope is slightly negative ($-0.015$~BPM~yr$^{-1}$, $R^{2}=0.076$), again barely perceptible across the study period. The nine-recording fast cluster, which includes Bylsma (1998), Feuermann (1937), both Fournier readings, Starker (1959), and Ma (1985), demonstrates that the high-energy reading of this movement has been continuously available throughout the recording era.

\subsection{Op.~102 No.~1, Adagio}

\begin{figure}[htbp]
    \centering
    \includegraphics[width=\columnwidth]{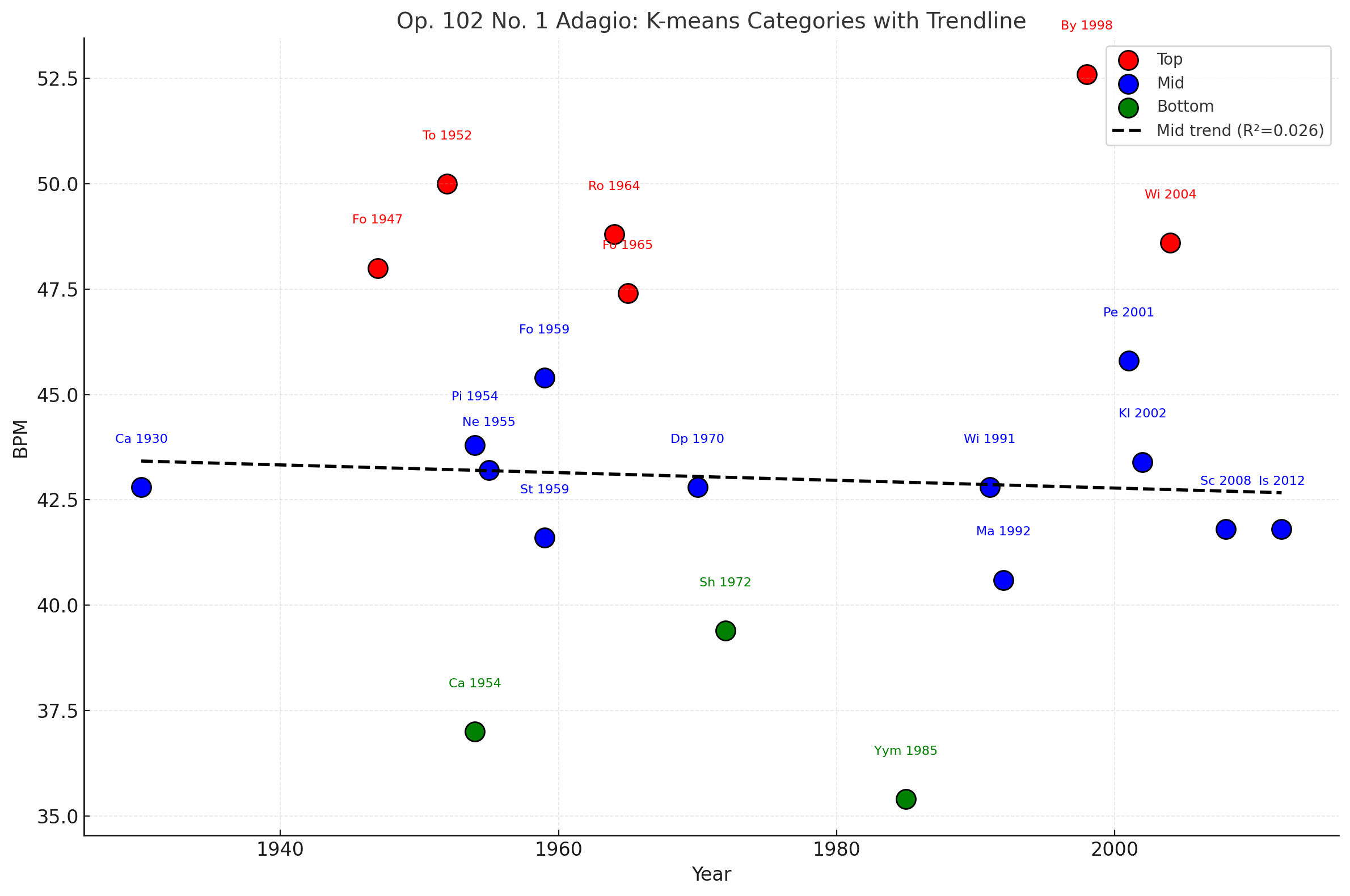}
    \caption{Scatter plot of average tempo by recording year for the Op.~102 No.~1 Adagio. Three clusters: slow (green, mean 37.3~BPM), mid-range (blue, mean 43.0~BPM), fast (red, mean 49.2~BPM). Mid-range regression $R^{2}=0.026$.}
    \label{fig:op102n1-adagio}
\end{figure}

Three clusters emerge cleanly for the Op.~102 No.~1 Adagio introduction. The slow cluster (mean 37.3~BPM) includes Casals (1954), Shafran (1972), and Ma (1985) --- three performances notable for their expansive phrasing and their chronological distribution across the full study period, demonstrating that the slow tradition was not a pre-war anomaly but a continuous alternative throughout the era. The mid-range cluster (mean 43.0~BPM, eleven recordings) is the most compact in the corpus, with a standard deviation under 1.5~BPM, indicating exceptional consensus within this tradition. The fast cluster (mean 49.2~BPM) comprises Fournier (1947, 1965), Rostropovich (1964), Tortelier (1952), Bylsma (1998), and Wispelwey (2004), cellists who prioritise rhythmic momentum in this ambiguous slow introduction.

The mid-range intra-cluster regression returns $R^{2}=0.026$, one of the lowest values in the corpus. This movement's central tradition has been not merely stable but essentially immobile across eight decades --- a finding that contrasts sharply with the Op.~102 No.~1 Allegro that follows it.

\subsection{Op.~102 No.~1, Allegro con brio}

\begin{figure}[htbp]
    \centering
    \includegraphics[width=\columnwidth]{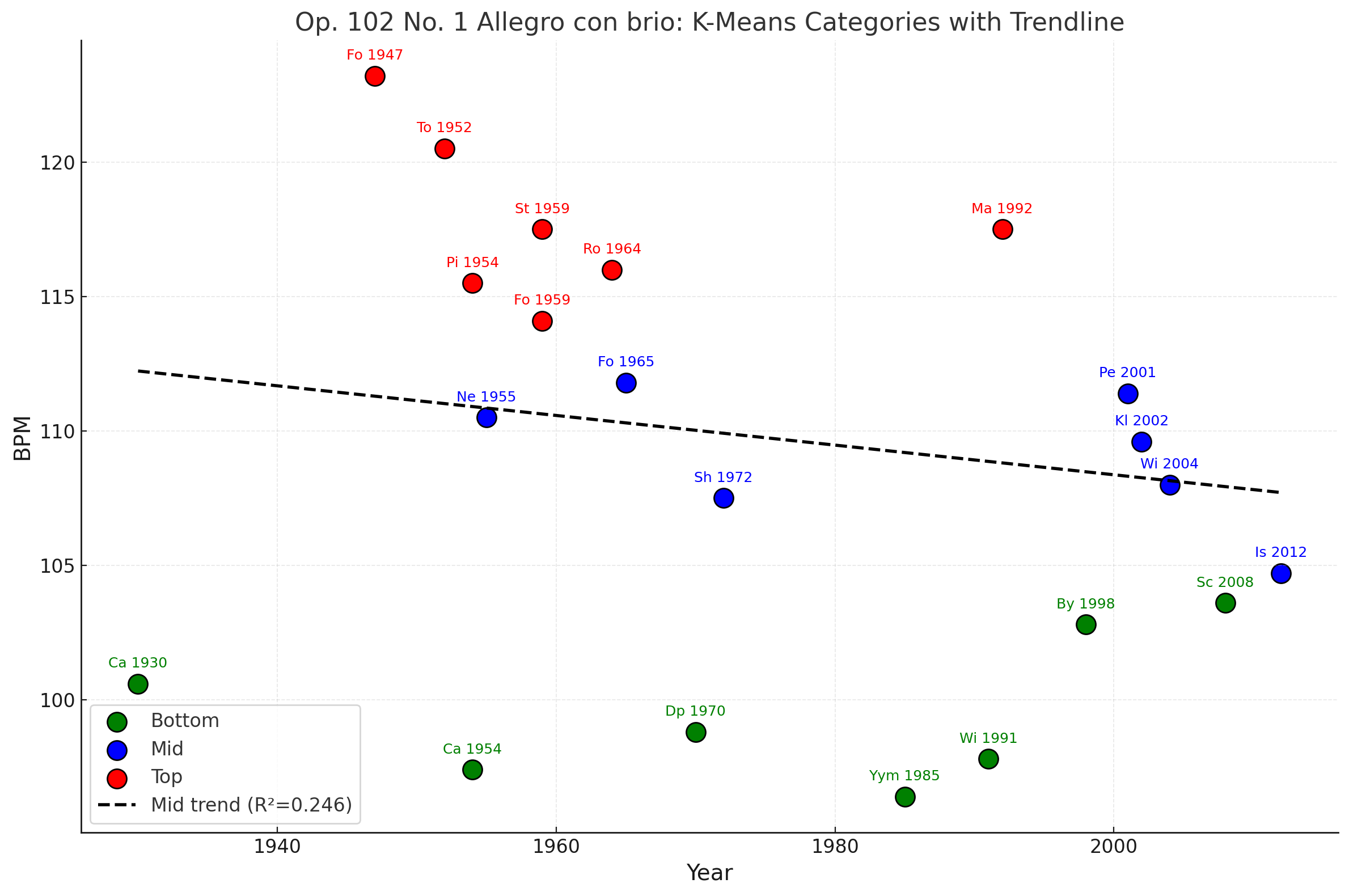}
    \caption{Scatter plot of average tempo by recording year for the Op.~102 No.~1 Allegro con brio. Three clusters: slow (green, mean 98.8~BPM), mid-range (blue, mean 110.5~BPM), fast (red, mean 121.4~BPM). Mid-range regression $R^{2}=0.246$, slope $= -0.032$~BPM~yr$^{-1}$ ($p=0.013$).}
    \label{fig:op102n1-allegro}
\end{figure}

The Allegro con brio of Op.~102 No.~1 is the most analytically complex movement in the corpus and the only one where intra-cluster regression yields a statistically significant result. Three clusters are clearly defined: slow (mean 98.8~BPM, six recordings including Casals 1954, du~Pr\'e 1970, Ma 1985, Wispelwey 1991, and Bylsma 1998), mid-range (mean 110.5~BPM, eight recordings from Feuermann 1937 through Wispelwey 2004), and fast (mean 121.4~BPM, seven recordings including Feuermann 1937, Fournier 1947, Rostropovich 1964, and Schott 2008).

The mid-range intra-cluster regression returns $R^{2}=0.246$ with a slope of $-0.032$~BPM~yr$^{-1}$ (slope $t$-statistic $= -2.67$, $p=0.013$), corresponding to a deceleration of approximately 3.2~BPM across the study period. Residual analysis reveals no systematic departures from linearity. This is the only movement in the corpus where the dominant tradition has undergone a statistically significant directional shift --- and notably, it is a deceleration rather than the acceleration that simple narratives of twentieth-century performance would predict.

The broad tempo spread across all three clusters (98.8--121.4~BPM, a range of 22.6~BPM between slow and fast centroids) is the widest in the corpus, reflecting the ambiguous character of this brief but intense Allegro in Beethoven's most formally experimental late sonata.

\subsection{Op.~102 No.~2, Adagio}

\begin{figure}[htbp]
    \centering
\includegraphics[width=\columnwidth]{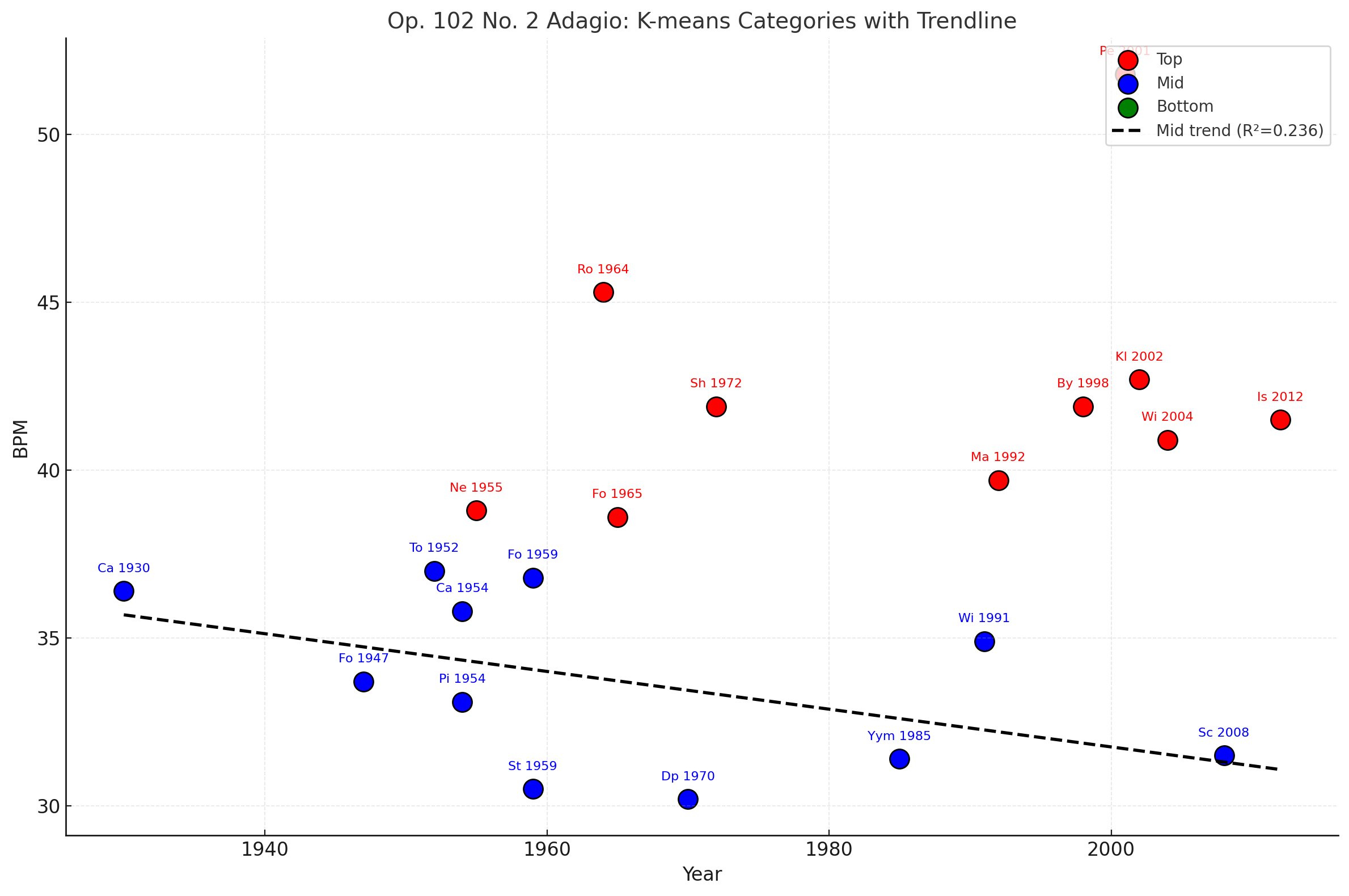}
    \caption{Scatter plot of average tempo by recording year for the Op.~102 No.~2 Adagio. Two populated clusters: mid-range (blue, mean 33.8~BPM) and fast (red, mean 42.3~BPM). Mid-range regression slope $= +0.008$~BPM~yr$^{-1}$ ($R^{2}=0.236$).}
    \label{fig:op102n2-adagio}
\end{figure}

The Op.~102 No.~2 Adagio yields two populated clusters and one empty slow category. The mid-range cluster (mean 33.8~BPM, fourteen recordings from Casals 1930 through Schott 2008) is the dominant tradition, with a positive intra-cluster regression slope ($+0.008$~BPM~yr$^{-1}$, $R^{2}=0.236$) indicating a modest but perceptible historical acceleration equivalent to approximately 0.8~BPM per decade. The fast cluster (mean 42.3~BPM, nine recordings from Nelsova 1955 through Isserlis 2012) comprises performances that pursue a markedly more driven pulse.

The contrast with the Op.~102 No.~1 Adagio is instructive. The No.~1 Adagio's mid-range cluster is essentially immobile ($R^{2}=0.026$); the No.~2 Adagio's mid-range cluster shows meaningful drift ($R^{2}=0.236$). This divergence is consistent with the different structural roles of the two slow movements: the No.~1 Adagio functions primarily as a transition preceding an energetic Allegro, while the No.~2 Adagio is the emotional centre of the sonata, a movement whose interpretive weight invites more varied re-evaluation across generations.

\subsection{Op.~102 No.~2, Allegro}

\begin{figure}[htbp]
    \centering
  \includegraphics[width=\columnwidth]{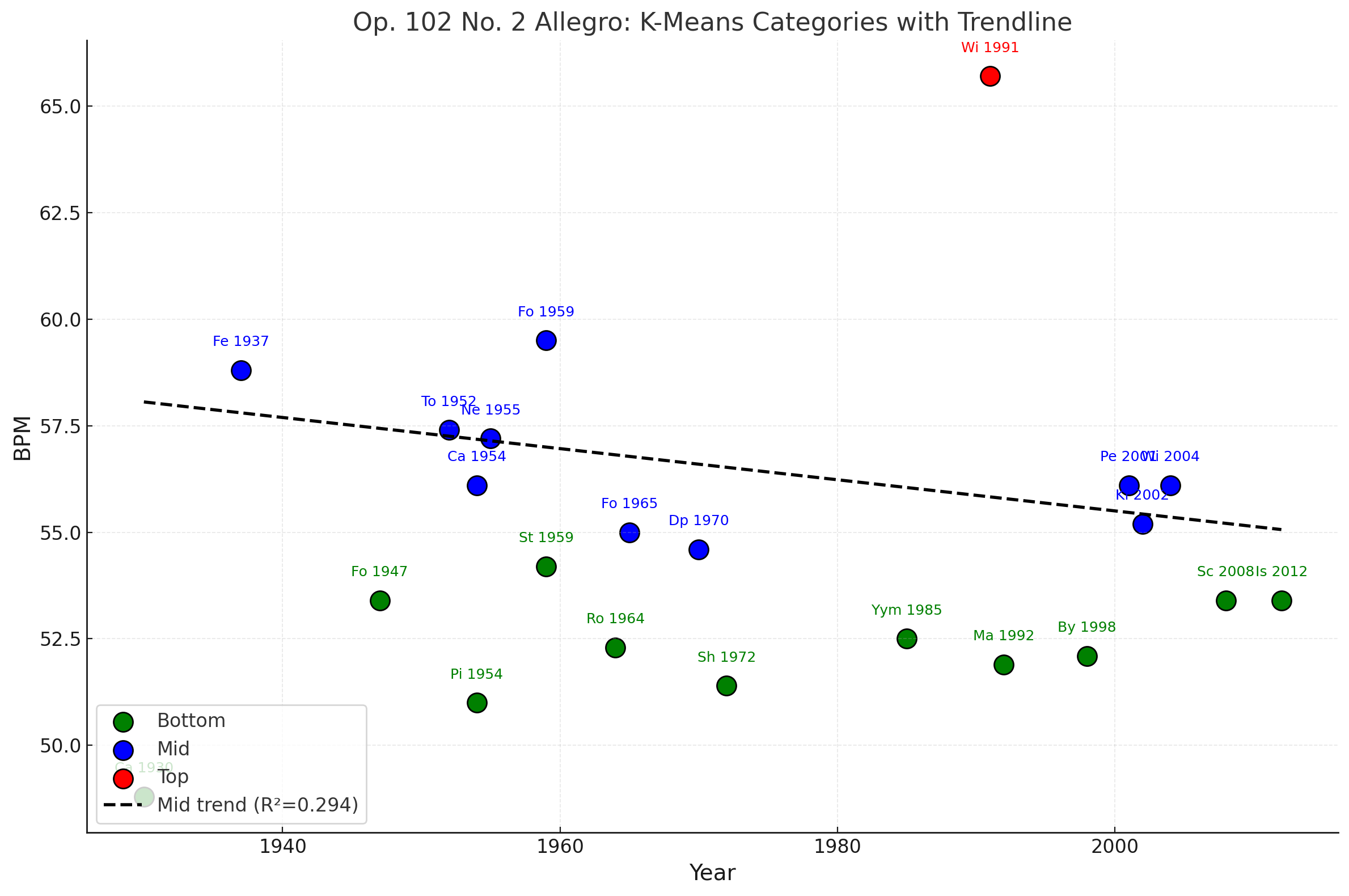}
    \caption{Scatter plot of average tempo by recording year for the Op.~102 No.~2 Allegro. Three clusters: slow (green, mean 52.2~BPM), mid-range (blue, mean 56.6~BPM), fast (red, single recording at 65.7~BPM). Mid-range regression slope $= +0.012$~BPM~yr$^{-1}$ ($R^{2}=0.08$).}
    \label{fig:op102n2-allegro}
\end{figure}

The Op.~102 No.~2 Allegro produces three clusters, but with an asymmetric distribution: slow (six recordings, mean 52.2~BPM), mid-range (fourteen recordings, mean 56.6~BPM, spanning 53.0--60.0~BPM), and fast (a single recording: Wispelwey 1991 at 65.7~BPM). The single-record fast cluster is an outlier in the strict statistical sense. The mid-range regression slope is positive and small ($+0.012$~BPM~yr$^{-1}$, $R^{2}=0.08$), suggesting a gradual but limited acceleration of 1.2~BPM per century. This contrasts with the Op.~102 No.~1 Allegro's deceleration, demonstrating that even adjacent movements in the same sonata can exhibit opposite directional tendencies within their dominant traditions.


\section{Synthesis}

\subsection{Cross-Movement Cluster Structure}

Table~\ref{table:cluster_summary_fast} and Table~\ref{table:cluster_summary_slow} summarise the cluster structures across all movements analysed. Several patterns emerge consistently.

\begin{table}[htbp]
\caption{Cluster summary: fast-character movements}
\centering
\small
\begin{tabular}{@{}llcccc@{}}
\toprule
\textbf{Movement} & \textbf{Cluster} & \textbf{N} & \textbf{$\bar{T}$} & \textbf{Range} & \boldmath{$R^{2}$} \\
\midrule
Op.\,5/1 Rondo   & Slow      & 3  & 78.0  & 75--82   & $\approx$0   \\
                  & Mid       & 13 & 83.1  & 80--86   & 0.001        \\
                  & Fast      & 4  & 90.2  & 88--92   & $\approx$0   \\
\addlinespace
Op.\,5/2 Rondo   & Mid       & 14 & 66.5  & 61--71   & 0.142        \\
                  & Fast      & 5  & 76.7  & 73--81   & ---          \\
                  & Slow      & 0  & ---   & ---      & ---          \\
\addlinespace
Op.\,69 Scherzo  & Mid       & 14 & 92.3  & 88--98   & 0.067        \\
                  & Fast      & 8  & 115.0 & 105--161 & ---          \\
                  & Slow      & 0  & ---   & ---      & ---          \\
\addlinespace
Op.\,69 Allegro  & Slow      & 6  & 145.0 & 142--148 & $\approx$0   \\
                  & Mid       & 7  & 148.6 & 147--151 & 0.076        \\
                  & Fast      & 9  & 156.0 & 152--161 & $\approx$0   \\
\addlinespace
Op.\,102/1 Allegro & Slow    & 6  & 98.8  & 97--103  & $\approx$0   \\
                  & Mid       & 8  & 110.5 & 110--118 & \textbf{0.246}\\
                  & Fast      & 7  & 121.4 & 116--161 & $\approx$0   \\
\addlinespace
Op.\,102/2 Allegro & Slow    & 6  & 52.2  & 30--57   & $\approx$0   \\
                  & Mid       & 14 & 56.6  & 53--60   & 0.080        \\
                  & Fast      & 1  & 65.7  & ---      & ---          \\
\bottomrule
\end{tabular}
\label{table:cluster_summary_fast}
\end{table}

\begin{table}[htbp]
\caption{Cluster summary: slow-character movements}
\centering
\small
\begin{tabular}{@{}llcccc@{}}
\toprule
\textbf{Movement} & \textbf{Cluster} & \textbf{N} & \textbf{$\bar{T}$} & \textbf{Range} & \boldmath{$R^{2}$} \\
\midrule
Op.\,69 Adagio   & Slow      & 6  & 36.3  & 31--39   & $\approx$0   \\
                  & Mid       & 10 & 42.7  & 42--46   & 0.012        \\
                  & Fast      & 4  & 48.3  & 48--53   & $\approx$0   \\
\addlinespace
Op.\,102/1 Adagio & Slow     & 3  & 37.3  & 36--39   & $\approx$0   \\
                  & Mid       & 11 & 43.0  & 42--46   & 0.026        \\
                  & Fast      & 6  & 49.2  & 48--53   & $\approx$0   \\
\addlinespace
Op.\,102/2 Adagio & Mid      & 14 & 33.8  & 30--37   & \textbf{0.236}\\
                  & Fast      & 9  & 42.3  & 39--53   & ---          \\
                  & Slow      & 0  & ---   & ---      & ---          \\
\bottomrule
\end{tabular}
\label{table:cluster_summary_slow}
\end{table}

\textit{Mid-range cluster dominance.} In every movement with at least two clusters, the mid-range cluster is the most populous, typically comprising 55--75\% of recordings. This consistency across eight movements and five sonatas confirms that there is always a normative interpretive consensus; a tempo that most performers converge on regardless of period, nationality, or pedagogical background. The regression-based average approximates this consensus but obscures the fact that a substantial minority of performers consistently diverges from it in both directions.

\textit{Absence of a slow cluster in fast-character movements.} The Op.~5 Rondo No.~2, Op.~69 Scherzo, and Op.~102 No.~2 Adagio all lack a slow cluster. These movements share a character that performers apparently treat as having a lower tempo bound: the Rondo of No.~2 with its relentless semiquaver activity, the Scherzo with its explosive articulation, and the No.~2 Adagio (paradoxically the most fluid of the three slow movements) with its searching harmonic language that resists the kind of structural suspension that a genuinely slow slow-movement tempo invites. This is a finding about the relationship between musical character and the range of tempo choices performers consider viable.

\textit{Intra-cluster stability as the dominant condition.} Of the eight movements examined, only one (Op.~102 No.~1 Allegro con brio) shows a statistically significant intra-cluster regression slope. The other seven mid-range clusters have $R^{2}$ values ranging from 0.001 to 0.142. This overwhelming internal stability is the paper's core empirical finding: tempo traditions in this repertoire are robust across generations, resisting the directional drift that would be expected if stylistic change were a uniform, field-wide process.

\subsection{Tempo, Duration, and the Aggregate Trend}

Table~\ref{table:tempo_duration} presents the aggregate tempo and duration changes from the 1930--1970 period to the 1970--2012 period for the second and third movements. Several movements show measurable aggregate acceleration alongside duration contraction. The Op.~102 No.~2 Adagio and Op.~69 Adagio cantabile show the largest tempo increases (+14.0\% and +13.9\% respectively), while the Op.~69 Scherzo shows a sharp aggregate deceleration ($-40.4\%$) primarily driven by the disappearance of the extreme early outliers (Casals 1930 at 161~BPM, Feuermann 1937 at 160~BPM) from later decades.

\begin{table}[htbp]
\caption{Aggregate tempo and duration changes by movement (1930--1970 vs.\ 1970--2012)}
\centering
\small
\begin{tabular}{@{}lcc@{}}
\toprule
\textbf{Movement} & \textbf{Tempo \%} & \textbf{Duration \%} \\
\midrule
Op.\,5/1 Rondo          & $+10.0$\% & $-9.1$\%  \\
Op.\,5/2 Rondo          & $+5.1$\%  & $-5.0$\%  \\
Op.\,69 Scherzo         & $-40.4$\% & $+67.9$\% \\
Op.\,69 Adagio cant.    & $+13.9$\% & $-12.5$\% \\
Op.\,69 Allegro vivace  & $+1.2$\%  & $+5.3$\%  \\
Op.\,102/1 Adagio       & $-2.3$\%  & $+2.1$\%  \\
Op.\,102/1 Allegro      & $+4.1$\%  & $-4.0$\%  \\
Op.\,102/2 Adagio       & $+14.0$\% & $-12.5$\% \\
Op.\,102/2 Allegro      & $+9.4$\%  & $-8.7$\%  \\
\bottomrule
\end{tabular}
\label{table:tempo_duration}
\end{table}

The clustering analysis reframes these aggregate figures. The Op.~69 Scherzo's $-40.4\%$ aggregate deceleration is not evidence of a historical retreat from fast tempi but of the disappearance of two exceptional outliers (Casals and Feuermann) from a fast cluster that otherwise persisted in modified form into later decades. Wispelwey (2004) and Maisky (1992) remain in the fast cluster; the extreme outer boundary of that cluster simply contracted as the specific historical conditions that produced Casals's and Feuermann's readings were not reproduced. Similarly, the modest positive tempo changes in Op.~5 Rondos and Op.~102 Allegros do not represent all performers uniformly accelerating; they represent a slight population shift toward the fast cluster in post-1970 decades, with the mid-range cluster itself remaining stable.

Duration changes correlate inversely with tempo changes (approximate correlation $|r| \approx 0.98$ across the movements in Table~\ref{table:tempo_duration}), confirming that global pacing choices largely govern total performance time and that the two measures are not independent. This near-perfect inverse correlation is itself evidence that the aggregate trends are real at the distributional level, even though they misrepresent the within-cluster dynamics.

\subsection{Performer Background and Cluster Membership}

A systematic examination of cluster membership against performers' generational cohort (pre-1930 born, 1930--1960 born, post-1960 born), national background (French, Russian, Spanish/Catalan, British, American, Hungarian, Dutch), and pedagogical lineage (teachers traceable through the Casals, Feuermann, Piatigorsky, and Rostropovich traditions) reveals no meaningful alignment between any of these biographical variables and cluster membership.

Performers sharing the same pedagogical lineage frequently appear in different clusters. Yo-Yo Ma (Rostropovich tradition) appears in the mid-range and fast clusters across different sonatas; Wispelwey appears in both mid-range and fast clusters within the same sonata's different movements. Cellists trained within the French tradition (Fournier, Tortelier) appear across all three clusters depending on the movement and the recording year. National schools, to the extent they can be identified from biographical data, do not systematically predict slower or faster readings.

This absence of correlation is itself a substantive finding. It means that the tempo tradition a performer adopts is an individual interpretive decision, not a culturally or pedagogically inherited one. The coexistence of traditions is not a matter of different schools maintaining their own norms in isolation; it is a matter of individual performers making independent choices within a shared field of possibilities.


\section{Discussion}

\subsection{The Ecological Model of Stylistic Change}

The results support and substantiate the ecological model proposed in Section~II. In ecology, the composition of a community at any moment reflects not which species have evolved but how the relative populations of coexisting species have shifted. The parallel in performance practice is precise. The slow, mid-range, and fast tempo traditions for each movement of Beethoven's piano and cello sonatas are not stages of evolution but coexisting approaches that have been simultaneously available to performers throughout the study period. What has changed is not the definition of these traditions (their BPM centroids are stable) but the number of performers who have chosen each one.

The ecological model accommodates all the findings of this paper more naturally than an evolutionary model does. The absence of a slow cluster in fast-character movements is explained not by the extinction of slow-tempo performers but by the structural constraints of those movements making extremely slow readings implausible to virtually all performers. The single case of significant intra-cluster drift (Op.~102 No.~1 Allegro) is explained not as evidence of uniform field-wide change but as the one movement in the corpus whose mid-range tradition has itself been subject to ongoing re-evaluation, perhaps because the movement's extreme character makes the question of an appropriate normative tempo genuinely unresolved.

This model has implications beyond this study. Any performance corpus in which a single regression line explains less than, say, 25\% of the total tempo variance is a candidate for clustering analysis. The low $R^{2}$ values of corpus-level regressions in Bowen~\cite{c1}, Cook~\cite{c2}, and Leech-Wilkinson~\cite{c3} suggest that this condition may be widespread. Clustering these corpora retrospectively would test whether the ecological model generalises beyond Beethoven chamber music.

\subsection{The Regression vs.\ Clustering Contrast}

The contrast between the single-regression and clustering accounts of the same data can be illustrated concretely using the Op.~5 No.~1 Rondo. A single regression of BPM against year across all twenty recordings returns a positive slope consistent with the general narrative of post-1970 acceleration. The $R^{2}$ of this regression is low, and the residuals are large and structurally patterned; systematically positive for the slow cluster recordings and systematically negative for the fast cluster recordings. The regression line passes through no natural grouping in the data; it is a line fitted to a trimodal distribution.

The clustering account disaggregates this picture. It shows three stable traditions (78.0, 83.1, 90.2~BPM) with near-zero within-cluster slopes. The aggregate positive trend is entirely explained by the fact that the three early Casals and du~Pr\'e recordings in the slow cluster are concentrated in the early decades: as these recordings recede into history and the post-1970 decades are filled by mid-range and fast performers, the distribution's weighted mean shifts upward slightly. No individual performer has changed their tempo; the population has changed.

This is a methodologically important distinction. It means that the standard narrative of twentieth-century tempo acceleration, insofar as it is based on regression analysis alone, may be fundamentally mischaracterising what is actually a change in the relative prevalence of coexisting traditions, not a change in the traditions themselves.

\subsection{Implications for Performance Practice Research}

The findings have three concrete implications for how empirical performance studies should be conducted and interpreted.

First, before fitting a regression line to corpus-level tempo data, researchers should apply cluster validity tests to determine whether the data has significant internal structure. If silhouette scores indicate multiple well-separated clusters, regression analysis of the full dataset is inappropriate as a primary analytical tool.

Second, cluster membership should be reported alongside regression results in any large-corpus tempo study. Even where the full-corpus regression is statistically significant, the within-cluster regressions tell a different and usually more informative story about the dynamics of stylistic change.

Third, the absence of a correlation between cluster membership and performers' biographical backgrounds is a finding that should be sought explicitly in any performance tradition study. Where such a correlation exists, it is evidence of school-level or cultural-level stylistic inheritance; where it is absent, as here, it is evidence that individual interpretive agency is the primary driver of stylistic diversity.


\section{Conclusion}

This paper has applied $k$-means clustering to bar-level tempo data from over one hundred movement-level recordings of Beethoven's five piano and cello sonatas, spanning 1930--2012, and has argued that the resulting cluster structure, typically three discrete traditions labelled slow, mid-range, and fast, is a more accurate and more informative description of the historical record than any single regression line can provide.

The key findings are: the mid-range cluster is dominant in every movement; slow clusters are absent from fast-character movements; intra-cluster regression slopes are negligible in seven of eight movements; the single statistically significant intra-cluster trend (Op.~102 No.~1 Allegro, $R^{2}=0.246$) describes a deceleration rather than the acceleration that simple historical narratives predict; and no correlation exists between cluster membership and performers' generational, national, or pedagogical backgrounds.

These findings support an ecological model of stylistic change in which coexisting traditions shift in relative prevalence over time, rather than an evolutionary model in which a single tradition gradually transforms. The aggregate tempo trends documented by earlier regression-based studies are real at the distributional level but misrepresent the within-tradition dynamics: what has changed is primarily the proportional weighting of stable traditions, not the traditions themselves.

The methodology presented here is transferable. Any large-corpus performance study with sufficient within-corpus variance (and low full-corpus $R^{2}$ values), is a candidate for this analysis. The ecological model of stylistic change, and the clustering methodology that reveals it, offers a new framework for understanding how performance practice evolves: not uniformly, not linearly, but as a shifting ecology of coexisting interpretive approaches.


\end{document}